\documentclass[twocolumn]{aastex63}

\newcommand{\kms}{km\,s$^{-1}$}

\usepackage{textcomp}
\usepackage{CJKutf8}

\shorttitle{A 99-minute WD+WD Binary from SDSS-V}
\shortauthors{Chandra et al.}

\graphicspath{{./}{fig/}}

\begin{document}

\title{A 99-minute Double-lined White Dwarf Binary from SDSS-V}

\correspondingauthor{Vedant Chandra}
\email{vedantchandra@g.harvard.edu}

\author[0000-0002-0572-8012]{Vedant~Chandra}
\affiliation{Department of Physics \& Astronomy, Johns Hopkins University, 3400 N Charles St, Baltimore, MD 21218, USA}
\affiliation{Center for Astrophysics $\mid$ Harvard \& Smithsonian, 60 Garden St, Cambridge, MA 02138, USA}

\author[0000-0003-4250-4437]{Hsiang-Chih~Hwang}
\affiliation{Department of Physics \& Astronomy, Johns Hopkins University, 3400 N Charles St, Baltimore, MD 21218, USA}

\author[0000-0001-6100-6869]{Nadia~L.~Zakamska}
\affiliation{Department of Physics \& Astronomy, Johns Hopkins University, 3400 N Charles St, Baltimore, MD 21218, USA}

\author[0000-0002-2761-3005]{Boris~T.~G{\"a}nsicke}
\affiliation{Department of Physics, University of Warwick, Coventry, CV4 7AL, UK}

\author[0000-0001-5941-2286]{J.~J.~Hermes}
\affiliation{Department of Astronomy, Boston University, 725 Commonwealth Ave., Boston, MA 02215, USA}

\author[0000-0003-3441-9355]{Axel~Schwope}
\affiliation{Leibniz-Institut f{\"u}r Astrophysik Potsdam (AIP), An der Sternwarte 16, 14482 Potsdam, Germany}

\author[0000-0003-3494-343X]{Carles~Badenes}
\affiliation{Department of Physics and Astronomy, University of Pittsburgh, 3941 O’Hara St, Pittsburgh, PA 15260, USA}

\author[0000-0002-2953-7528]{Gagik~Tovmassian}
\affiliation{Instituto de Astronomia, Universidad Nacional Autonoma de Mexico, Apdo. Postal 877, Ensenada, Baja California 22800, Mexico}

\author[0000-0002-4791-6724]{Evan~B.~Bauer}
\affiliation{Center for Astrophysics $\mid$ Harvard \& Smithsonian, 60 Garden St, Cambridge, MA 02138, USA}

\author[0000-0002-6579-0483]{Dan~Maoz}
\affiliation{School of Physics and Astronomy, Tel-Aviv University, Tel-Aviv 6997801, Israel}

\author[0000-0003-3903-8009]{Matthias~R.~Schreiber}
\affiliation{Departamento de Física, Universidad Técina Federico Santa María, Avenida España 1680, Valparaíso, Chile}
\affiliation{Millennium Nucleus for Planet Formation (NPF), Valparaíso, Chile}

\author[0000-0002-2398-719X]{Odette~F.~Toloza}
\affiliation{Department of Physics, University of Warwick, Coventry, CV4 7AL, UK}
\affiliation{Departamento de Física, Universidad Técina Federico Santa María, Avenida España 1680, Valparaíso, Chile}
\affiliation{Millennium Nucleus for Planet Formation (NPF), Valparaíso, Chile}

\author[0000-0002-2200-2416]{Keith~P.~Inight}
\affiliation{Department of Physics, University of Warwick, Coventry, CV4 7AL, UK}

\author[0000-0003-4996-9069]{Hans-Walter~Rix}
\affiliation{Max-Planck-Institut f{\"u}r Astronomie, K{\"o}nigstuhl 17, D-69117 Heidelberg, Germany}

\author[0000-0002-4462-2341]{Warren~R.~Brown}
\affiliation{Smithsonian Astrophysical Observatory, 60 Garden Street, Cambridge, MA 02138 USA}

\begin{abstract}

\noindent %
We report the discovery of SDSS J133725.26+395237.7 (hereafter SDSS\,J1337+3952), a double-lined white dwarf (WD+WD) binary identified in early data from the fifth generation Sloan Digital Sky Survey (SDSS-V). The double-lined nature of the system enables us to fully determine its orbital and stellar parameters with follow-up Gemini spectroscopy and {\it Swift} UVOT ultraviolet fluxes. The system is nearby ($d = 113$ pc), and consists of a $0.51\, M_\odot$ primary and a $0.32\, M_\odot$ secondary. SDSS\,J1337+3952 is a powerful source of gravitational waves in the millihertz regime, and will be detectable by future space-based interferometers. Due to this gravitational wave emission, the binary orbit will shrink down to the point of interaction in $\approx 220$ Myr. The inferred stellar masses indicate that SDSS\,J1337+3952 will likely not explode as a Type Ia supernova (SN\,Ia). Instead, the system will probably merge and evolve into a rapidly rotating helium star, and could produce an under-luminous thermonuclear supernova along the way. The continuing search for similar systems in SDSS-V will grow the statistical sample of double-degenerate binaries across parameter space, constraining models of binary evolution and SNe\,Ia. 

\end{abstract}


\section{Introduction}

Compact binaries that contain white dwarfs (WDs), neutron stars, and/or black holes are at the core of many long-standing puzzles of modern astrophysics. Double-degenerate (WD+WD) binaries are of particular importance, in part because they may be a major source of Type Ia supernovae (SNeIa; see \citealt{Maoz2014} for a review), cosmological standard candles used to measure the accelerating expansion of the Universe \citep[e.g.,][]{Riess1998,Perlmutter1999}. Double-degenerate binaries will also be the largest population of gravitational wave sources detectable by future space-based observatories \citep[e.g.,][]{Marsh2011a,Kupfer2018,Lamberts2019,Li2020}. Studying the population of double-degenerates across a wide range of periods and masses contributes to our understanding of binary evolution from common envelopes to mergers \citep[e.g.,][]{Nelemans2000,Maxted2002, Marsh2004,VanDerSluys2006,Webbink2008,Brown2016a,Inight2021}. 

Most detached double-degenerate binaries are discovered by measuring changes in the radial velocities (RVs) of photospheric absorption lines over time \citep[e.g.,][]{Brown2010,Napiwotzki2020a}. Usually, one WD dominates the flux contribution in the spectrum, producing one set of absorption lines that vary over time due to Doppler shifts \citep[e.g.,][]{Brown2011}. In a small fraction of double-degenerates, the component WDs have comparable flux contributions, producing a double-lined (SB2) system. Double-lined binaries are particularly useful since measurements of both stars' velocities constrain the entire orbital solution of the system, most crucially the stellar masses. However, only $\approx 20$ double-lined double-degenerate binaries with well-measured parameters are currently known \citep[e.g.,][]{Saffer1988,Marsh1995b,Moran1997,Parsons2011,Marsh2011, Kilic2020,Parsons2020,Kilic2021b}. 

The fifth-generation Sloan Digital Sky Survey (SDSS-V; \citealt{Kollmeier2017}) is the first all-sky spectroscopic survey to explicitly target WDs with minimal well-measured selection effects. Identifying and characterizing double-degenerate binaries is a core goal of the SDSS-V Milky Way Mapper science program. Each co-added SDSS-V spectrum is composed of numerous 15-minute sub-exposures taken consecutively or split between different nights. These sub-exposures can be used to search for RV variability and identify binary systems \citep[e.g.,][]{Badenes2009,Schwope2009,Breedt2017}. This method is most sensitive to short orbital periods $\lesssim 1$~day, since this increases the probability that successive spectra will detect RV shifts. Due to the medium resolution of the spectra, the method is also most sensitive to high-amplitude RV variations, i.e. systems with extreme mass ratios. SDSS-V has already observed $\approx$ 50,000 sub-exposures of $\approx$ 6,000 unique white dwarfs, with the eventual goal of observing $\simeq$ 100,000 white dwarfs identified from \textit{Gaia} data \citep{Fusillo2019,Fusillo2021}. 

Here we report the discovery of SDSS J133725.26+395237.7 (hereafter SDSS\,J1337+3952), a 99-minute double-lined binary composed of two hydrogen atmosphere (DA) WDs. We identified SDSS\,J1337+3952 during a systematic search for RV-variable systems in the first year of SDSS-V. We jointly analyze the SDSS-V data with follow-up time-resolved spectroscopy and broadband photometry to fully determine the orbital and stellar parameters of this system. We summarize our observations in $\S$\ref{sec:obs} and describe our model-fitting analysis in $\S$\ref{sec:analysis}. We present our results in $\S$\ref{sec:results}, and discuss the past and future evolution of the system in $\S$\ref{sec:discuss}.

\section{Observations}\label{sec:obs}

\subsection{SDSS-V}\label{sec:obs.sdss}

\begin{figure}
    \centering
    \includegraphics[width=\columnwidth]{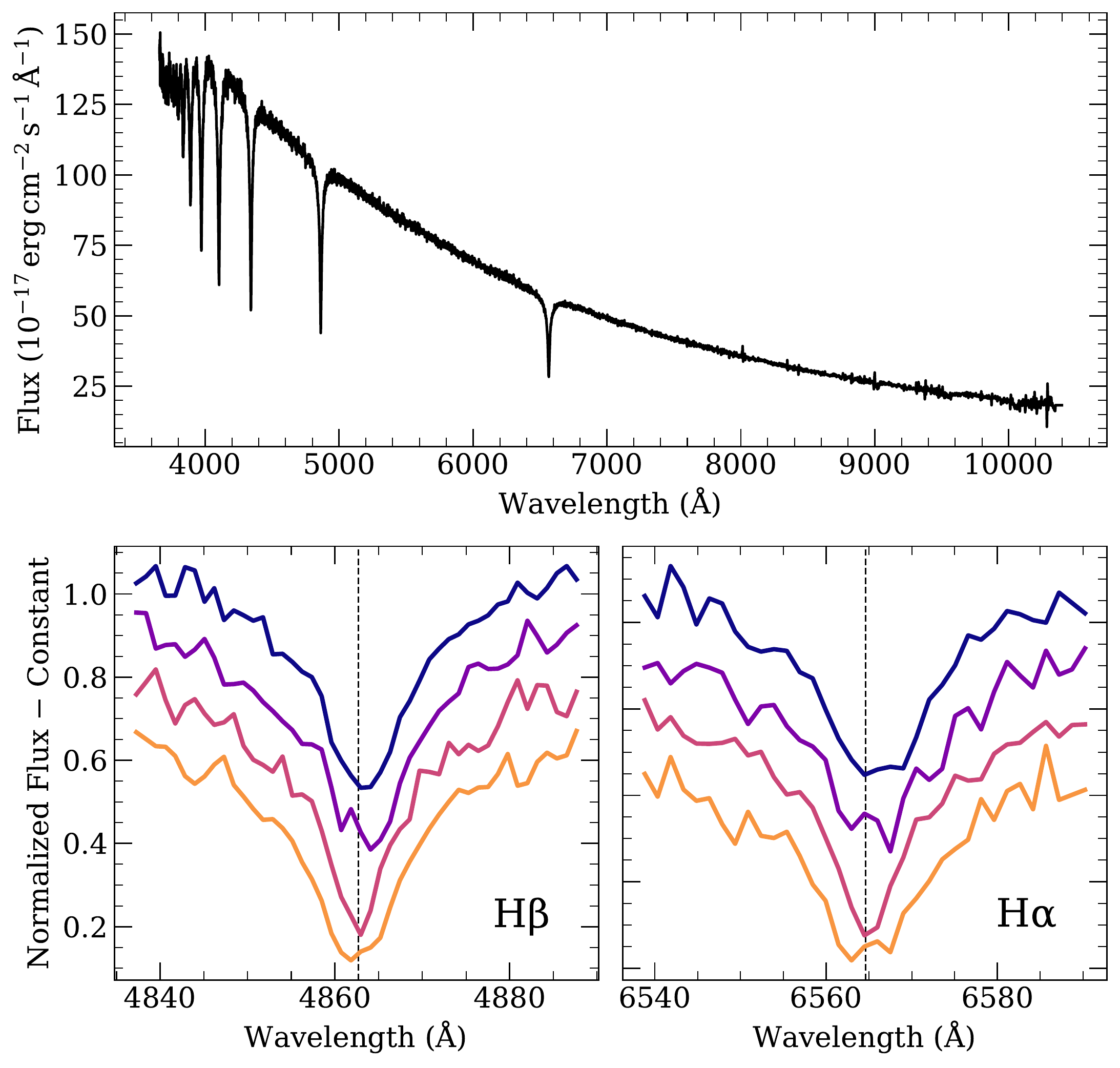}
    \caption{Discovery spectra of SDSS\,J1337+3952. Top: Stacked SDSS-V spectrum derived by median-combining all 19 sub-exposures from several nights. Bottom: Continuum-normalized Balmer lines of four consecutive sub-exposures from 2021~March~20, spanning $\sim 1$ hour in total. The exposures are vertically offset for clarity, with the first exposure at the top. The absorption lines visibly split in phase with each other, suggesting that the system is an unresolved binary with two Doppler-shifted components.}
    \label{fig:sdss_summary}
\end{figure}

SDSS\,J1337+3952 was observed by the Baryon Oscillation Spectroscopic Survey spectrograph (BOSS; \citealt{Smee2013}) as a part of the fifth generation Sloan Digital Sky Survey (SDSS-V; \citealt{Kollmeier2017}). It was originally selected as a WD target from the catalog of \cite{Fusillo2019}. SDSS\,J1337+3952 has a \textit{Gaia} Early Data Release 3 (EDR3) prior-informed distance $d = 113.3 \pm 0.5$\,pc and tangential velocity $43 \pm 1$~\kms{}, implying a thin-disk origin for the system \citep{GaiaCollaboration2018b,GaiaCollaboration2021,Lindegren2021,Bailer-Jones2020}. Between 2021~March~20 and 2021~July~4, SDSS\,J1337+3952 was observed with BOSS for a total of 19 sub-exposures. Each exposure was 900\,s long and covered a wavelength range of 3600--$10\,000$\,\AA{} with $R \approx 1800$ resolution. At first glance, the optical spectrum of SDSS\,J1337+3952 is typical for a DA WD, with strong hydrogen Balmer absorption lines being the only discernible features (Figure \ref{fig:sdss_summary}, top). However, several sub-exposures show splitting in the H$\alpha$ and H$\beta$ absorption lines, identifying this system as a double-lined binary candidate (Figure \ref{fig:sdss_summary}, bottom). Furthermore, SDSS\,J1337+3952 sits a magnitude brighter than the cooling track for $\sim 0.6\,M_\odot$ WDs on the {\em Gaia} color-magnitude diagram, suggesting that the total flux is a composite of two unresolved stars. 

\subsection{Gemini GMOS}\label{sec:obs.gmini}

We observed SDSS\,J1337+3952 with the Gemini North Multi-Object Spectrograph (GMOS-N; \citealt{Hook2004}) as a part of GN-2021A-FT-112 (PI: Chandra). We used the high-resolution R831 grating centered at 575.0\,nm with a 0.5\arcsec\ slit, for a resolving power $R \approx 4400$ across $450-700$\,nm. We obtained a run of 10 consecutive exposures on 2021~June~2, and 19 exposures on 2021~June~7, with individual exposure times of 300\,s throughout. The June 7 exposures were split into two runs (10 and nine exposures, respectively) separated by three hours to increase the time baseline. We obtained CuAr arc exposures before each run to ensure a precise wavelength calibration. We reduced our data using the PypeIt utility \citep{pypeit:joss_pub,pypeit:zenodo}. This included bias-correction, flat-fielding, wavelength calibration, and source extraction. We performed a second-order flexure correction to each exposure's wavelength solution using night sky lines. The Gemini spectra covered the H$\alpha$ and H$\beta$ Balmer lines, but the H$\alpha$ line had twice the signal-to-noise ratio (S/N). The single-exposure H$\alpha$ spectra have $\text{S/N} \approx 10-25$ per pixel depending on the observing run. Furthermore, the H$\alpha$ line is ideal for measuring RVs of WDs since it is minimally affected by asymmetric broadening \citep{Halenka2015}. Therefore, we only use the Gemini H$\alpha$ spectra in our RV analysis. 

\subsection{Spectral Energy Distribution}\label{sec:obs.sed}

\begin{deluxetable}{lcr}\label{tab:sed}
\tablewidth{\columnwidth}
\tablecaption{Adopted spectral energy distribution of SDSS\,J1337+3952}
\tablecolumns{3}
\tablehead{
\colhead{Band} & \colhead{\hspace{1.25cm}$\lambda_{\text{ref}}$ (\AA{})} \hspace{1.25cm} & \colhead{AB Magnitude}}
\startdata
$uvw2$ & 2055 & 18.40 $\pm$ 0.05 \\
$uvm2$ & 2246 & 17.94 $\pm$ 0.04 \\
$uvw1$ & 2580 & 17.70 $\pm$ 0.03 \\
$U$ & 3467 & 17.14 $\pm$ 0.03 \\
$u$ & 3557 & 17.14 $\pm$ 0.03 \\
$g$ & 4702 & 16.65 $\pm$ 0.03 \\
$r$ & 6175 & 16.65 $\pm$ 0.03 \\
$i$ & 7489 & 16.70 $\pm$ 0.03 \\
$z$ & 8947 & 16.82 $\pm$ 0.03 \\
$J$ & 12358 & 17.18 $\pm$ 0.07 \\
$H$ & 16457 & 17.77 $\pm$ 0.20 \\
$Ks$ & 21603 & 18.01 $\pm$ 0.25 \\
$W1$ & 33526 & 18.95 $\pm$ 0.05 \\
$W2$ & 46028 & 19.84 $\pm$ 0.19 \\
\enddata
\tablecomments{$\lambda_{\text{ref}}$ is the approximate pivot wavelength of the photometric band. Conversions from Vega to AB magnitudes were performed using coefficients from \cite{Blanton2007} for 2MASS and \citep{Cutri2015} for \textit{WISE}.}
\end{deluxetable}

We assembled the spectral energy distribution (SED) of SDSS\,J1337+3592 using VizieR \citep{vizier}. SDSS\,J1337+3592 has secure archival photometry in the Sloan $ugriz$ \citep{Fukugita1996,Gunn1998,Doi2010,Blanton2017,Ahumada2020}, 2MASS $J,H,Ks$ \citep{Skrutskie2006}, and \textit{WISE} $W1,W2$ \citep{Wright2010} bands. Since SDSS\,1337+3592 is nearby and lies well out of the Galactic plane ($b \approx 74$ degrees), we assume negligible interstellar extinction. This assumption is supported by the three-dimensional dust maps of \cite{Green2018}. 

We observed SDSS\,J1337+3592 with the Ultraviolet and Optical Telescope (UVOT; \citealt{Roming2005}) on the Neils Gehrels \textit{Swift} space observatory \citep{Gehrels2004} as a target of opportunity between 2021~June~22--25 (Target ID 14380, PI: Tovmassian). We obtained a total of 982\,s of exposure in the 1928\,\AA{} \textit{uvw2} band, 736\,s in the 2246\,\AA{} \textit{uvm2} band, 1216\,s in the 2600\,\AA{} \textit{uvw1} band, and 962\,s in the 3465\,\AA{} \textit{U} band. We performed photometry with a 5\arcsec\ extraction aperture and a background region a few arcseconds south of the target using the Web--HERA tool \citep{Pence2012} provided by The High Energy Astrophysics Science Archive Research Center (HEASARC). We used HEASOFT v6.28 software and calibration procedures to determine the {\sl UV} magnitudes and fluxes \citep{Breeveld2011}. As expected for a detached binary, no X-ray emission was detected at the source position by the \textit{Swift} X-Ray Telescope (XRT; \citealt{Burrows2005}), with an upper limit of $0.0026$ count/s in the entire $0.3-10$\,keV range.

We summarize our assembled spectral energy distribution in Table~\ref{tab:sed}. To prevent any photometric bands from dominating our fitting procedure due to underestimated systematics, we adopt a floor uncertainty of 0.03 mag in all bands \citep[e.g.,][]{Bergeron1997}. 

\subsection{TESS Light Curve}

Some SB2 WD systems exhibit eclipses, enabling their orbital periods and scaled radii to be precisely measured \citep[e.g.,][]{Parsons2011}. SDSS\,J1337+3592 was observed by the {\em Transiting Exoplanet Survey Satellite} ({\em TESS}; \citealt{Ricker2015}) at a 2-minute cadence for nearly one month in Sector\,23 (targeted as TIC\,22846882). We downloaded and processed the \textit{TESS} light curve with \texttt{lightkurve} \citep{LightkurveCollaboration2018,Ginsburg2019}. We corrected the data to remove long-term systematic trends, and searched for coherent variability with a Lomb-Scargle periodogram \citep{Lomb1976,Scargle1982}. We fail to detect any periodic signals in the {\em TESS} light curve above the 1\% level, after adjusting for crowding of the source in the extracted {\em TESS} aperture. The non-detection of eclipses suggests that the system's inclination does not have an edge-on configuration. This is confirmed by our model fitting in $\S$\ref{sec:analysis.phot}, where we find that the inclination of the system is low enough to explain the non-detection of eclipses. 

\section{Analysis}\label{sec:analysis}

In this section, we describe our two-stage approach to characterize SDSS\,J1337+3952. First, we fit the time-resolved Gemini H$\alpha$ spectra to determine the binary orbital parameters of the system. Next, we simultaneously fit the SED, continuum-normalized Balmer lines, and Keplerian constraints to derive the system inclination and stellar parameters of both component WDs. 

\subsection{Orbital Parameters from Time-Resolved Spectra}\label{sec:analysis.gemspec}

\begin{figure*}
    \centering
    \includegraphics[height=0.35\textwidth]{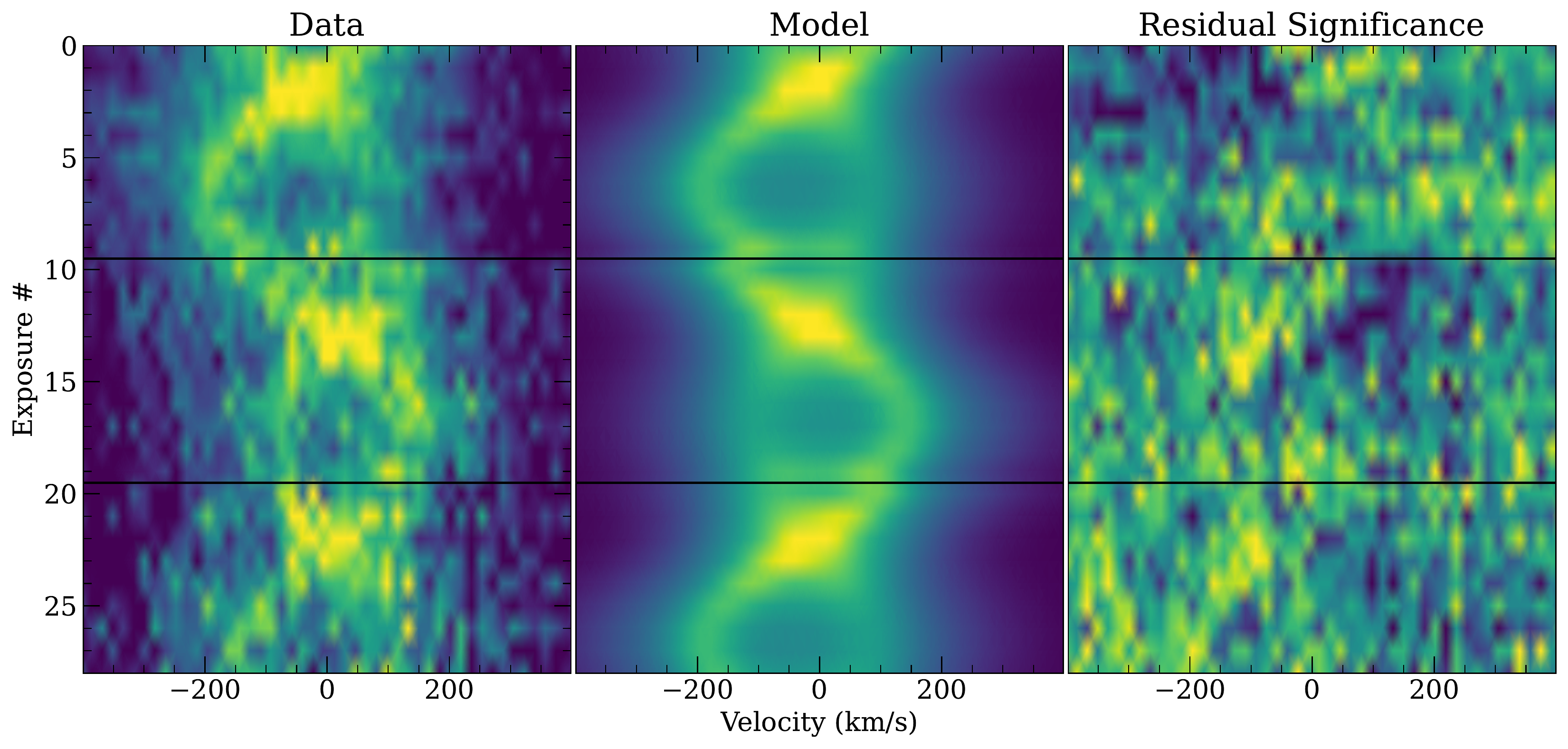}
    \includegraphics[height=0.35\textwidth]{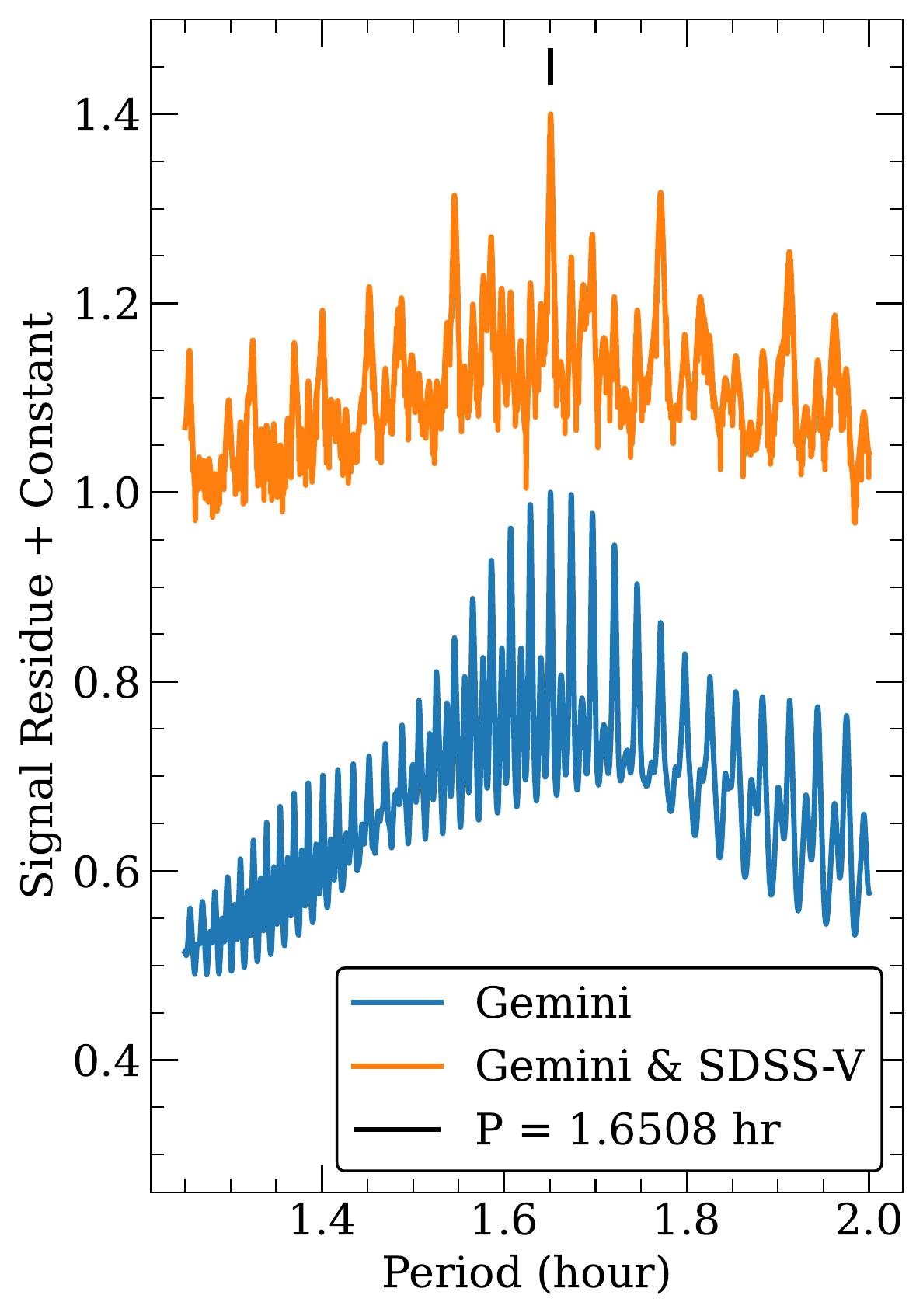}
    \caption{Left: Continuum-normalized Balmer H$\alpha$ spectra of SDSS\,J1337+3952 from Gemini GMOS across three runs, separated by black lines, along with the double-lined binary model. The residual significance is the difference between the data and model divided by the data uncertainty. The left and center colormaps have the same scale ($0.55-1$ in normalized flux), and the right colormap is scaled to saturate at 3-sigma. The spectrogram plots are smoothed via linear interpolation \citep{Gouraud1971}. Right: $\chi^2$ periodogram of the H$\alpha$ model as a function of period, keeping all other parameters fixed to their final values. The blue periodogram is the signal residue of the Gemini spectra alone, and the orange curve is the average of the Gemini and SDSS-V periodograms, vertically offset by a constant for visual clarity.}
    \label{fig:gemini_spec}
\end{figure*}

To determine the orbital parameters of SDSS\,J1337+3952, we fit the time-resolved Gemini H$\alpha$ spectra with a double-lined binary model. Due to the medium S/N of the spectra, and the fact that the component stars' absorption lines overlap in a number of exposures, we do not measure radial velocities from individual exposures. Rather, we fit the line profile and orbital model to all 29 exposures simultaneously. We follow the convention of denoting the more massive component as the `primary.'

We model the orbit with velocity semi-amplitudes $K_i$ (where $i$ indexes the stellar components), period $P$, and zero point orbital phase in hours since the first exposure $\phi$. We allow each star to have its own systemic velocity $\gamma_i$, because WDs with different masses will have different systemic velocities due to the gravitational redshift effect \citep[e.g.,][]{Einstein1916,Falcon2010,Chandra2020b}. Close WD binaries are usually assumed to have circularized orbits due to drag forces during their past common envelope evolution \citep{Paczynski1976}. To test this assumption, we first fit our data allowing for eccentric orbits. The model fit is not improved when eccentricity is included, and we can reject eccentricity $e \gtrsim 0.06$ at the 3-sigma level. We therefore assume a circular orbit with zero eccentricity in our subsequent analysis. 

We model each star's contribution to the continuum-normalized H$\alpha$ spectrum as the sum of two Gaussians with a shared centroid but independent width $\sigma$ and amplitude $a$. This produces four free shape parameters per star $\sigma_{i,j}$, $a_{i,j}$, where $i$ indexes the star and $j$ indexes the star's pair of Gaussian profiles. The Gaussian centroids are set to the star's modelled orbital velocity at each exposure's phase, and then the profiles from both stars are added to produce the double-lined model. The exposure time of each spectrum spans only $5\%$ of an orbital phase, so we assume the intra-exposure RV `smearing' of the spectrum is negligible at our resolution and S/N. In total, there are six orbital parameters and eight Gaussian shape parameters, for a total of 14 free parameters in our H$\alpha$ model. 

We continuum-normalize the Gemini H$\alpha$ spectra by dividing out a straight line fitted $400-500$\,\kms{} away from the theoretical rest-frame wavelength. We compute a $\chi^2$ residual between the model and observed spectra across all 29 exposures at once (Figure~\ref{fig:gemini_spec}, left). We proceed with two-step fitting approach. In the first step, we minimize the $\chi^2$ residual over all parameters using the nonlinear least-squares utility \texttt{lmfit} \citep{Newville2014}. In the second step, we derive a periodogram by computing $\chi^2$ over a finely-spaced grid of periods (keeping all other parameters fixed) and selecting the period with the lowest $\chi^2$. We iterate over both steps, using the periodogram period to initialize the least-squares fit, until the period and $\chi^2$ converge. Finally, using this solution as an initial point, we sample the posterior distributions of all parameters using the affine-invariant Markov Chain Monte Carlo (MCMC) sampler \texttt{emcee} \citep{Foreman-Mackey2019}. We select the MCMC sample with the lowest $\chi^2$ as our best-fitting parameter set, and derive uncertainties by computing the standard deviation of the MCMC chains. Our fitted parameters and their uncertainties are summarized in Table $\ref{tab:params}$. For brevity we omit the eight Gaussian shape parameters, which vary between $45\,\text{kms}^{-1} \lesssim \sigma_{i,j} \lesssim 155\,\text{kms}^{-1}$ and $0.1 \lesssim a_{ij} \lesssim 0.2$ (see Figure \ref{fig:halpha_corner} for all posterior parameter distributions). The best-fit period is $P = 1.65082(09)$\,hr, almost exactly 99 minutes.

To confirm that our Gemini-inferred period is robust against aliases, we repeat our H$\alpha$ analysis on the lower-resolution SDSS-V data. We select 10 SDSS-V sub-exposures that were all coincidentally taken within a week of our Gemini observing run. We do not use the entire set of 19 SDSS-V exposures since their long time baseline introduces significant fine-structure aliasing into the periodogram. Keeping the orbital parameters fixed to their Gemini-inferred values, we fit an independent set of Gaussian shape parameters to the SDSS-V spectra. We then compute the $\chi^2$ statistic of the SDSS-V spectra over a grid of periods, keeping all other parameters fixed. The best-fitting period from the SDSS-V data alone is within $\approx 3$ seconds of the Gemini-inferred period. We normalize the Gemini and SDSS-V periodograms to the $[0,1]$ range by computing their signal residue, $SR(P) = \min[\chi^2(P)] / \chi^2(P)$ \citep{Hippke2019}. By averaging these periodograms, we verify that the inclusion of the SDSS-V data promotes the correct peak while damping the aliases (Figure~\ref{fig:gemini_spec}, right). 

\subsection{Stellar Parameters from Spectrophotometry}\label{sec:analysis.phot}

We determine the stellar parameters of both WDs in SDSS\,J1337+3952 by simultaneously fitting the parallax, SED, and continuum-normalized Balmer lines on the SDSS-V spectrum. The free parameters are the effective temperatures $T_{\text{eff,i}}$ and surface gravities $\log{g}_i$ of both stars, and the inclination of the system $i$. Our likelihood function treats the photometric datapoints, spectroscopic datapoints, and measured velocity semi-amplitudes from $\S$\ref{sec:analysis.gemspec} as Gaussian random variables. Each parameter set $\left\{T_{\text{eff,i}},\log{g}_i,i\right\}$ produces a synthetic dataset of these observables whose likelihood can be calculated according to the chi-square distribution. In the following paragraphs we detail each component of our likelihood function. 

\begin{figure}
    \centering
    \includegraphics[width=1\columnwidth]{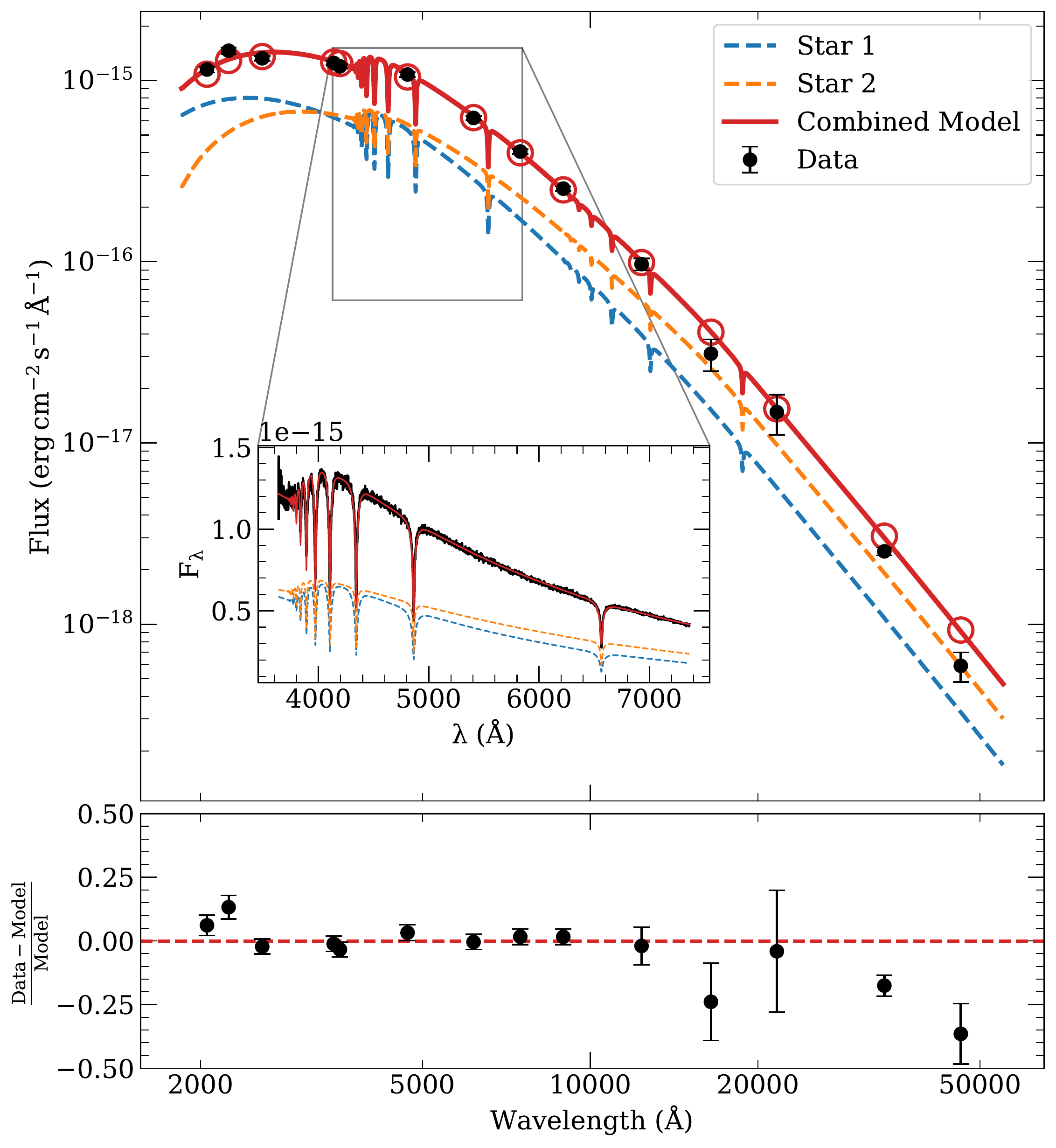}
    \includegraphics[width=1\columnwidth]{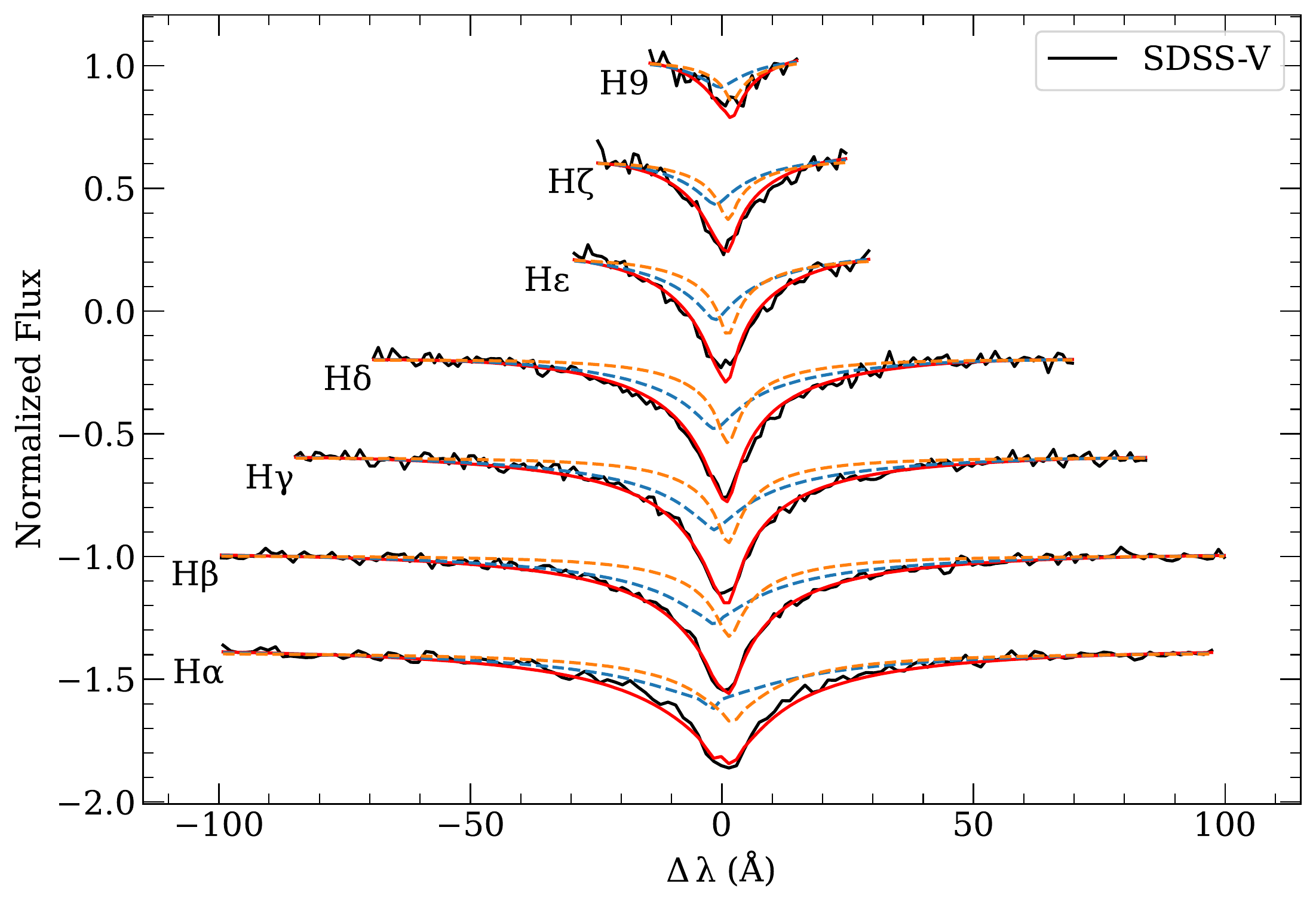}
    \caption{Simultaneous fit to the SED and Balmer absorption lines of SDSS\,J1337+3952. Top: The observed spectral energy distribution from UV--NIR with 1-sigma error bars. We show each star's contribution to the total model, and the composite model's integrated photometry is overlaid with red circles. The inset compares the composite model to the averaged SDSS-V spectrum. We use a multiplicative fourth-order Chebyshev polynomial to correct the spectrum's absolute flux calibration. Bottom: Continuum-normalized Balmer lines on the SDSS-V spectrum averaged from four sub-exposures at similar orbital phase. The Balmer lines are vertically offset for clarity and arranged H$9$--H$\alpha$ from top to bottom. The dashed curves approximately indicate each star's contribution at that orbital phase.}
    \label{fig:sed_spec_fit}
\end{figure}

From each parameter set of $T_{\text{eff,i}}$ and $\log{g}_i$, we compute the stellar masses $M_i$ and radii $R_i$ using theoretical WD sequences. For the more massive primary, we interpolate thick hydrogen-atmosphere WD evolutionary sequences that assume a carbon/oxygen (C/O) core composition\footnote{\url{https://www.astro.umontreal.ca/~bergeron/CoolingModels/}} \citep{Kowalski2006,Tremblay2011,Bedard2020}. For the low-mass secondary, we interpolate helium (He) core sequences from \citet{Istrate2016} that include the effects of element diffusion, with an assumed progenitor metallicity of $Z=0.02$. We compute each star's theoretical spectrum by interpolating a grid of 1D hydrogen-atmosphere model spectra\footnote{\url{http://svo2.cab.inta-csic.es/theory/newov2/index.php?models=koester2}} \citep{Koester2010}. We scale the model spectra to the respective stellar radii and parallax-inferred distance, and then add the fluxes. We integrate the model spectra under the relevant transmission curves to derive a synthetic SED \citep{pyphot2020}. We compute a photometric chi-square likelihood $\mathcal{L}_{\text{phot}}$ comparing the synthetic and observed SED in the UVOT, Sloan, 2MASS, and \textit{WISE} bands (Figure~\ref{fig:sed_spec_fit}, top). 

To fit the continuum-normalized hydrogen Balmer lines, we select and median-combine four SDSS-V sub-exposures in which the velocity difference of the component stars is maximal. According to the ephemeris derived in $\S$\ref{sec:analysis.gemspec}, these exposures are within $20\%$ of one another in orbital phase. By co-adding these spectra, we achieve a higher S/N at the cost of some RV `smearing' in the line core, which is minor compared to the width of the absorption lines. The Balmer lines of synthetic DA spectra have well-known biases due to limitations of the mixing-length approximation for convection \citep[e.g.,][]{Tremblay2010}. In particular, 1D models predict systematically higher $\log{g}$ at $T_{\text{eff}} \lesssim 14\,000$\,K than full 3D models. To account for this, we invert the 1D $\to$ 3D parameter corrections defined in \cite{Tremblay2013} to interpolate the appropriate 1D spectra from \cite{Koester2010} for a given set of sampled `3D' $T_{\text{eff,i}}$ and $\log{g}_i$. We Doppler shift the model spectra to their appropriate radial velocities at the chosen orbital phase and convolve them to the BOSS resolution and sampling. We continuum-normalize the Balmer lines from H$\alpha$-H$9$, and compute a spectroscopic chi-square likelihood $\mathcal{L}_{\text{spec}}$ comparing the composite model spectrum to the observation (Figure~\ref{fig:sed_spec_fit}, bottom). 

Finally, we include Keplerian constraints from the orbital solution derived in $\S$\ref{sec:analysis.gemspec}. For a given orbital period $P$, each parameter set uniquely predicts velocity semi-amplitudes $K_i$ via Kepler's law,%
\begin{equation}\label{kep1}
    K_1 = \left( \frac{2 \pi G \sin^3{i} \left(M_1 + M_2 \right)}{P \left( 1 + M_1/M_2\right)^3} \right)^{1/3}
\end{equation} %
\begin{equation}\label{kep2}
    K_2 = K_1\frac{M_1}{M_2}
\end{equation}%
where $G$ is Newton's gravitational constant. For a given set of stellar parameters and inclination, we compute the velocity semi-amplitudes predicted by Equations~\ref{kep1}-\ref{kep2}, fixing the orbital period $P$ to its adopted value from $\S$\ref{sec:analysis.gemspec}. We compare these predicted velocities $K_i$ to their observed values from $\S$\ref{sec:analysis.gemspec} with a chi-square Keplerian likelihood $\mathcal{L}_{\text{Kep}}$.

 Our combined likelihood for each parameter set $\left\{T_{\text{eff,i}},\log{g}_i,i\right\}$ is $\mathcal{L} = \mathcal{L}_{\text{phot}} \times \mathcal{L}_{\text{spec}} \times \mathcal{L}_{\text{Kep}}$. The photometric likelihood constrains the stellar temperatures via the shape of the SED, as well as the stellar radii via the total light emitted at a certain parallax-inferred distance. The SDSS-V spectroscopic likelihood constrains the effective temperatures and surface gravities of both stars since the Balmer lines are sensitive to pressure broadening \citep[e.g.,][]{Tremblay2009}. The Keplerian likelihood constrains both the total mass and mass ratio of the system. This latter constraint is unique to double-lined binary systems, and is crucial to break the degeneracy between the $\log{g}_i$ of the component stars. 
 
For numerical stability, we compute and add the three log-likelihoods (0.5 times the respective $\chi^2$ statistics). We maximize the likelihood $\mathcal{L}$ with nonlinear least-squares \citep{Newville2014,Virtanen2020}, and then explore the posterior parameter distributions with \texttt{emcee}. To propagate the distance uncertainty into our parameter uncertainties, we sample and marginalize over the distance as well, with a strong prior set by the \textit{Gaia} EDR3 measurement \citep{Bailer-Jones2020}. We select the MCMC sample with the lowest $\chi^2$ as our best-fit parameter set, and derive uncertainties by computing the standard deviation of the MCMC chains. We illustrate the posterior parameter distributions in Figure \ref{fig:phot_corner}. 
 
 Figure~\ref{fig:sed_spec_fit} compares our best-fit stellar model to the observed broadband photometry and SDSS-V spectrum. The stellar parameters and their uncertainties are summarized in Table~\ref{tab:params}. Uncertainties on all derived parameters like $M_i$ and $R_i$ are propagated via random sampling, with the mean and standard deviation of $10^6$ Monte Carlo samples reported. We emphasize that we report formal statistical uncertainties only. Systematic uncertainties in DA white dwarf parameters can be around 2\% in $T_{\text{eff}}$ and 0.1 dex in $\log{g}$ \citep{Tremblay2010, Tremblay2019}. Additionally, the choice of core composition and physics in the adopted evolutionary sequences can introduce a further systematic uncertainty in the derived masses of order $\approx 0.02\,M_\odot$. We have verified that our assumption of an He core for the secondary is well-founded; if we re-fit our data with C/O cores assumed for both stars, the derived secondary mass is $M_2 \approx 0.28\,M_\odot$, well within the regime in which He core models are more appropriate. 

\begin{deluxetable}{lr}\label{tab:params}
\tablewidth{\columnwidth}
\tablecaption{Adopted measurements of SDSS\,J1337+3952}
\tablehead{
\colhead{Parameter} & \colhead{Value}}
\startdata
\sidehead{\textit{Gaia} EDR3}
Source ID & 1500004000845782912 \\
RA (degrees) & 204.35524 \\
Dec. (degrees) & 39.87712 \\
$G$ (mag) & 16.59 \\
$G_{\mathrm{BP}} - G_{\mathrm{RP}}$ (mag) & 0.31 \\
$\varpi$ (mas) & 8.80 $\pm$ 0.04 \\
$d$ (pc, \citealt{Bailer-Jones2020}) & 113.3 $\pm$ 0.5 \\ 
$\mu$ (mas/yr) & 80.59 $\pm$ 0.04 \\
\tableline
\sidehead{Gemini H$\alpha$ Fit}
$\gamma_1$ (km\,s$^{-1}$) & -8 $\pm$ 2 \\
$\gamma_1 - \gamma_2$ (km\,s$^{-1}$) & 11 $\pm$ 3 \\
$K_1$ (km\,s$^{-1}$) & 100 $\pm$ 4 \\
$K_2$ (km\,s$^{-1}$) & 168 $\pm$ 3 \\
$P$ (hour) & 1.65082 $\pm$ 0.00009 \\
$\phi$ (hour) & 0.660 $\pm$ 0.004 \\
\tableline
\sidehead{SED + Balmer Fit}
$T_{\text{eff},1}$ (K) & 9390 $\pm$ 60 \\
$\log{g}_1$ (dex) & 7.85 $\pm$ 0.03 \\
$T_{\text{eff},2}$ (K) & 7940 $\pm$ 70 \\
$\log{g}_2$ (dex) & 7.32 $\pm$ 0.02 \\
$i$ (degrees) & 34 $\pm$ 1 \\
\tableline
\sidehead{Derived Parameters}
$M_1$ ($M_\odot$) & 0.51 $\pm$ 0.01 \\
$M_2$ ($M_\odot$) & 0.32 $\pm$ 0.01 \\
$R_1$ ($R_\odot$) & 0.0141 $\pm$ 0.0002 \\
$R_2$ ($R_\odot$) & 0.0204 $\pm$ 0.0002 \\
$\tau_{c,1}$ (Myr) & $\approx 600$ \\
$\tau_{c,2}$ (Myr) & $\approx 1200$ \\
$v_{g,1} - v_{g,2}$ (km\,s$^{-1}$) & 13 $\pm$ 1 \\
$\tau_{\text{GW}}$ (Myr) & $\approx 220$ \\
$\mathcal{A}/10^{-22}$ (dimensionless) & 4.4 $\pm$ 0.1 \\
\enddata
\tablecomments{We report formal statistical uncertainties only. Uncertainties on derived parameters are propagated via Monte Carlo sampling, with the standard deviation of $10^6$ samples reported.}
\end{deluxetable}

As a check on our stellar parameters, we test whether the gravitational redshifts predicted by our stellar model are consistent with the systemic velocity difference measured with the Gemini H$\alpha$ spectra in $\S$\ref{sec:analysis.gemspec}. For a star with a given mass $M$ and radius $R$, the gravitational redshift of light leaving the stellar photosphere is given by $v_g = {GM}/{Rc}$, where $c$ is the speed of light \citep{Einstein1916}. Substituting our stellar parameters into this relation, we predict a difference in gravitational redshifts $v_{g,1} - v_{g,2} = 13 \pm 1$ \kms{}. This is consistent with our measured difference in systemic velocities $\gamma_{1} - \gamma_{2} = 11 \pm 3$ \kms{}, securing our confidence in the adopted stellar parameters. 

\section{Results}\label{sec:results}

\begin{figure}
    \centering
    \includegraphics[width=\columnwidth]{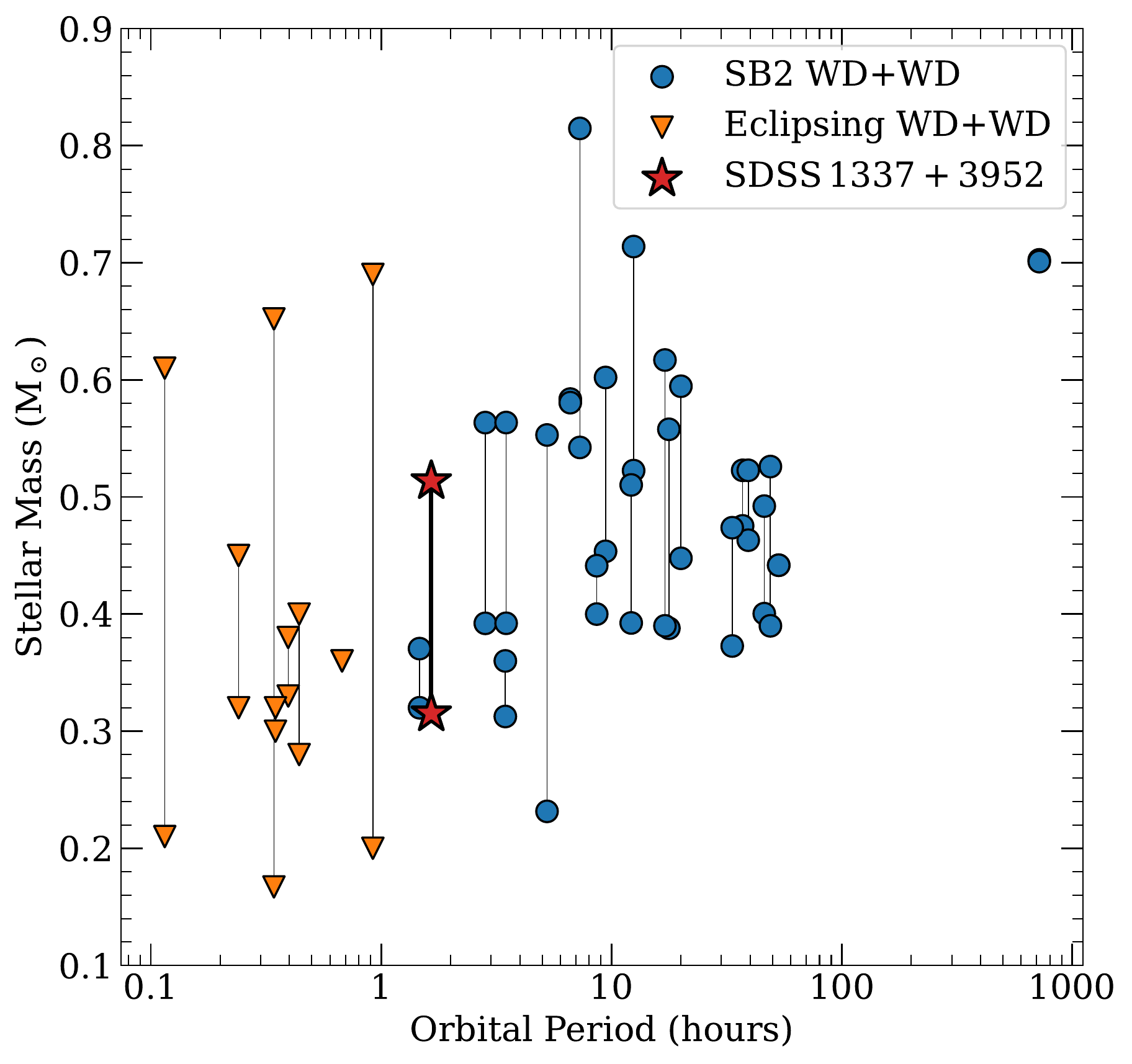}
    \caption{Contextualizing the mass and period measurements of SDSS\,J1337+3952 within a sample of known double-degenerate WD binaries. SDSS\,J1337+3952 is shown as a red star. In blue we show the sample of all known double-lined (SB2) double-degenerates compiled by \cite{Kilic2021b}. In orange we show a sample of eclipsing double-degenerates discovered by \cite{Burdge2020}. The primary and secondary masses for each system are joined with a line. Characteristic uncertainties are smaller than the marker sizes.}
    \label{fig:mass_period}
\end{figure}

We have presented the discovery and analysis of SDSS\,J1337+3952, a double-lined WD binary with a 99-minute orbital period. In Figure~\ref{fig:mass_period} we compare the mass and period measurements of SDSS\,J1337+3952 to the broader sample of all known double-lined double-degenerate binaries with well-constrained parameters \citep{Kilic2021b}. SDSS\,J1337+3952 is one of a few double-lined systems containing a WD in the extremely-low-mass (ELM; \citealt{Brown2010}) regime $\lesssim 0.3\,M_\odot$ \citep{Parsons2011,Bours2014}. The rest of the ELM sample is mostly composed of single-lined systems in which one star dominates the flux contribution \citep{Brown2020}. We also overlay in Figure~\ref{fig:mass_period} a sample of eclipsing binaries found by \cite{Burdge2020}. Since low-mass WDs are larger and more luminous due to the WD mass--radius relation, the eclipsing search method is biased towards very short-period systems containing one or two low-mass WDs. A magnitude-limited spectroscopic search like SDSS-V will also be biased to find a higher fraction of luminous low-mass binaries, since they are detectable out to a larger search volume. 

Since white dwarfs gradually lose heat via radiation after they form, their present-day stellar parameters can be used to estimate a `cooling age' since their formation \citep[e.g.,][]{Fontaine2002}. Interpolating our best-fit stellar parameters for SDSS\,J1337+3952 from Table~\ref{tab:params} onto theoretical evolutionary sequences, we derive respective cooling ages of $\tau_{c,1} \approx 600\,\text{Myr}$ (using C/O core sequences from \citealt{Bedard2020}) and $\tau_{c,2} \approx 1200\,\text{Myr}$ (using He core sequences from \citealt{Istrate2016}). The primary could also plausibly be an He core WD, but its parameters lie off the model grid computed by \cite{Istrate2016}. Therefore, its cooling age should be viewed with some caution. 

\begin{figure}
    \centering
    \includegraphics[width=\columnwidth]{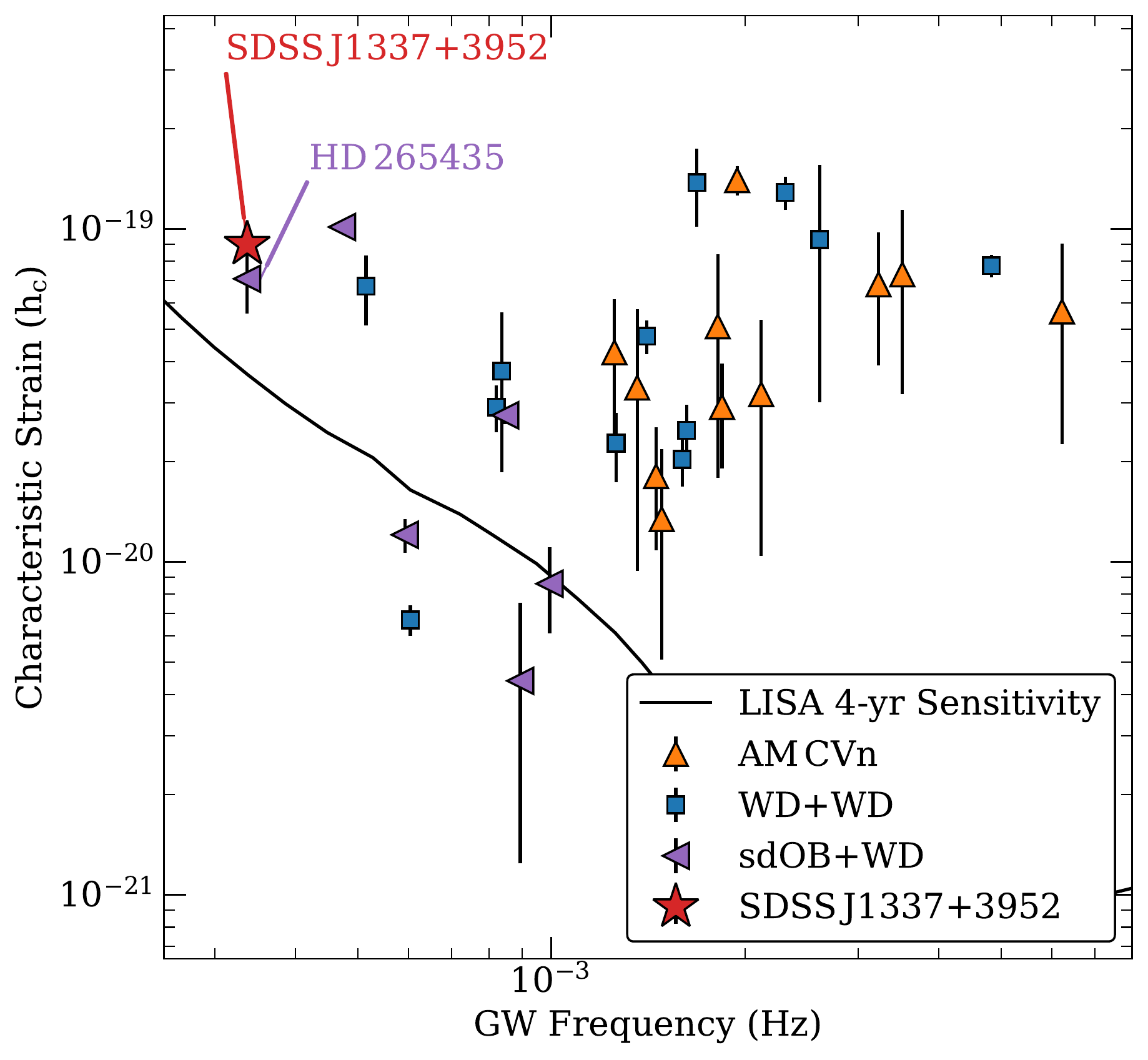}
    \caption{GW characteristic strain ($h_c = \mathcal{A}\sqrt{2 T_{\text{obs}} / P}$) against GW frequency ($f_{\text{GW}} = 2/P$), assuming an observation time $T_{\text{obs}}$ of 4 years. We display a sample of known {\em LISA} sources including detached double WDs, AM Canum Venaticorum (AM\,CVn) systems, and subdwarf-WD (sdOB+WD) binaries \citep{Kupfer2018,Burdge2020,Pelisoli2021}. We overlay the approximate 4-yr {\em LISA} sensitivity curve from \cite{Robson2019}. Horizontal errors are negligible at this scale, and vertical error bars are 1-sigma after propagating mass and distance uncertainties for each system. The vertical error for SDSS\,J1337+3952 is smaller than the marker size, and the nearby error bar is from the datapoint below it.}
    \label{fig:gw_strain}
\end{figure}

Due to its proximity to Earth and short period, SDSS\,J1337+3952 is among the strongest known sources of gravitational waves (GWs) in the mHz frequency regime (Figure~\ref{fig:gw_strain}). Using formulae from \cite{Kupfer2018} and propagating uncertainties via Monte Carlo sampling, we derive a dimensionless GW strain amplitude $\mathcal{A} = (4.4 \pm 0.1) \times 10^{-22}$. The eventual S/N of the gravitational wave signal is dependent on other factors like sky location, orbital inclination, and detector effects. The relatively low inclination of SDSS\,J1337+3952 favors its S/N for a future space-based mission like the {\em Laser Interferometer Space Antenna} ({\em LISA}; \citealt{Amaro-Seoane2017}), since it promotes a strong signal in both the plus and cross polarizations \citep[e.g.,][]{Shah2012,Shah2013}. Following the methodology outlined in \cite{Robson2019}, we estimate SDSS\,J1337+3952 will reach S/N $\approx 5$ over a nominal four-year {\em LISA} mission. SDSS\,J1337+3952 could be a useful verification system due its precise estimate for the GW amplitude. This precision stems from its double-lined nature --- which allows both component masses to be measured --- and its accurate parallax-inferred distance from \textit{Gaia}. 

Looking ahead, SDSS\,J1337+3952's orbit will continuously shrink as it loses energy via GW emission. For a system with a period $P$ in hours and stellar masses in $M_\odot$, the merging timescale due to gravitational wave emission \citep[e.g.,][]{Landau1975} is given by %
\begin{equation}\label{tau_gw}
    \tau_{\text{GW}} = 10 \cdot \frac{\left( M_1 + M_2 \right)^{1/3}}{M_1 M_2} P^{8/3}\ \text{Myr}
\end{equation}%
Substituting our best-fit parameters for SDSS\,J1337+3952, we derive $\tau_{\text{GW}} \approx 220$\,Myr. Therefore, SDSS\,J1337+3952 joins a small class of detached double-degenerate systems whose orbits will shrink to the point of interaction well within a Hubble time. Assuming that the cooling age of the younger WD corresponds to the time since the most recent common envelope (CE) phase ($\approx 600$\,Myr), we can invert Equation~\ref{tau_gw} to derive an initial post-CE orbital period $\approx 150$\,minutes, which is consistent with past surveys of post-CE binaries \citep{Moran2011}. From this we infer that SDSS\,J1337+3952's period has already reduced by $\approx 30\%$ due to GW emission since the double-degenerate binary was formed. 

\section{Discussion}\label{sec:discuss}

Binary star interactions are required to produce all known low-mass WDs, since the Universe is not old enough to evolve isolated stars to WD masses $\lesssim 0.45\, M_\odot$ \citep[e.g.,][]{Iben1990,Marsh1995}. The standard formation scenario for close double WDs is two consecutive CE phases, during which dynamically unstable mass transfer leads to the formation of a gaseous envelope engulfing both stars \citep[e.g.,][]{Webbink1984}. However, the formation of some close double He core WDs seems to be inexplicable assuming two CE phases \citep{Nelemansetal2000} and in these cases it appears more likely that the first phase of mass transfer was stable and non-conservative \citep{Webbink2008,Woods2012}. Given that the low-mass secondary has a larger cooling age, a plausible formation scenario is outlined by \cite{Woods2012}. According to this scenario, the present-day secondary was initially the more massive star, and it ascended the giant branch first and stably lost mass to the present-day primary. The subsequent giant-branch evolution of the present-day primary would have created a common envelope, leading to energy loss and in-spiral, eventually creating a detached double-degenerate binary with a period of a few hours. The progenitor stellar masses were likely between $\sim 1-1.5\,M_\odot$ \citep[e.g.,][]{Li2019}, with mass being lost from the system both during the initial stable mass transfer and when the common envelope was ejected.

In $\approx 220~\text{Myr}$ when the WDs in SDSS\,J1337+3952 are close enough to interact, mass transfer will ensue from the secondary WD onto the primary WD. The mass ratio $q = M_2/M_1 = 0.62 \pm 0.02$ is almost large enough for dynamically unstable mass transfer to be guaranteed \citep{Marsh2004}. The precise fate of the system depends on the spin-orbit coupling and core composition of the accreting primary. If mass transfer is stable, the system could form a binary of the AM Canum Venaticorum (AM\,CVn) class \citep[e.g.,][]{Nelemans2001,Ramsay2018}. It is then plausible that accreted helium will undergo successive shell flashes, culminating in an under-luminous thermonuclear supernova of Type .Ia \citep[][]{Bildsten2007,Shen2009,Shen2010}. However, it is unlikely that a system with such a moderate mass ratio could sustain stable mass transfer, since the accretion will be via direct impact rather than via an accretion disk. Further, CE dynamical friction could push nearly all close double-degenerate systems to eventually merge \citep{Shen2015,Brown2016a}.

Assuming the more likely scenario that mass transfer is eventually unstable, SDSS\,J1337+3952 will probably merge to form a rapidly rotating helium star which will end its life as a helium-atmosphere (DB) WD \citep{Saio1998,Saio2000,Saio2002,Schwab2018}. It may experience an intermediate evolutionary phase as an R Coronae Borealis (R\,Cr\,B) class star if the primary has a carbon-oxygen core \citep[e.g.,][]{Clayton2011,Zhang2014,Schwab2019}. During the accretion and merger, the primary is unlikely to become massive enough for the core to detonate as a Type Ia supernova and unbind the star \citep{Shen2014, Yungelson2017a}. However, if the secondary is sufficiently helium-rich, the violent merger could still detonate helium and produce an under-luminous SN\,.Ia without requiring stable mass transfer \citep{Guillochon2010,Pakmor2013,Shen2014b}.

SDSS\,J1337+3952 is presumably the first of many double-degenerate binaries that will be revealed by SDSS-V. By refining our search routines and applying them to the full survey data that will be obtained over the next few years, we expect to discover dozens more single-lined and double-lined WD+WD systems. We are continuing a systematic search and follow-up program for WDs with significant RV variations in SDSS-V. Although any spectroscopic survey will ultimately be magnitude-limited, a concerted effort is being made in SDSS-V to target WDs all across the color-magnitude diagram. As we have shown here, multi-epoch SDSS-V spectra can reveal binary candidates that warrant follow-up observations, and can also themselves be utilized to constrain a system's stellar and orbital parameters. In the near future, detailed analyses of individual systems will improve our understanding of stellar evolution and binary interaction. Statistical studies of the growing sample of double-degenerates will provide a cross-sectional perspective into the formation, evolution, and fate of compact binaries.

\newpage

\begin{acknowledgments}

We are grateful to the anonymous referee for their positive and constructive report. VC thanks Jennifer Andrews and Thomas Seccull for assistance with GMOS observations, and J. Xavier Prochaska, Shenli Tang, and Ryan Cooke for assistance with GMOS data reduction. VC and HCH. were supported in part by Space\,@\,Hopkins. VC, HCH, and NLZ were supported in part by NASA-ADAP 80NSSC19K0581. GT acknowledges support from PAPIIT project IN110619. MRS acknowledges support from Fondecyt (grant 1181404) and ANID -- Millennium Science Initiative Program -- NCN19\_171. OFT was supported by a Leverhulme Trust Research Project Grant and FONDECYT project 32103.

Funding for the Sloan Digital Sky Survey V has been provided by the Alfred P. Sloan Foundation, the Heising-Simons Foundation, and the Participating Institutions. SDSS acknowledges support and resources from the Center for High-Performance Computing at the University of Utah. The SDSS web site is \url{www.sdss5.org}. SDSS is managed by the Astrophysical Research Consortium for the Participating Institutions of the SDSS Collaboration, including the Carnegie Institution for Science, Chilean National Time Allocation Committee (CNTAC) ratified researchers, the Gotham Participation Group, Harvard University, The Johns Hopkins University, L'Ecole polytechnique f\'{e}d\'{e}rale de Lausanne (EPFL), Leibniz-Institut f\"{u}r Astrophysik Potsdam (AIP), Max-Planck-Institut f\"{u}r Astronomie (MPIA Heidelberg), Max-Planck-Institut f\"{u}r Extraterrestrische Physik (MPE), Nanjing University, National Astronomical Observatories of China (NAOC), New Mexico State University, The Ohio State University, Pennsylvania State University, Smithsonian Astrophysical Observatory, Space Telescope Science Institute (STScI), the Stellar Astrophysics Participation Group, Universidad Nacional Aut\'{o}noma de M\'{e}xico, University of Arizona, University of Colorado Boulder, University of Illinois at Urbana-Champaign, University of Toronto, University of Utah, University of Virginia, Yale University, and Yunnan University. 

Based in part on observations obtained at the international Gemini Observatory, a program of NSF’s NOIRLab, which is managed by the Association of Universities for Research in Astronomy (AURA) under a cooperative agreement with the National Science Foundation on behalf of the Gemini Observatory partnership: the National Science Foundation (United States), National Research Council (Canada), Agencia Nacional de Investigaci\'{o}n y Desarrollo (Chile), Ministerio de Ciencia, Tecnolog\'{i}a e Innovaci\'{o}n (Argentina), Minist\'{e}rio da Ci\^{e}ncia, Tecnologia, Inova\c{c}\~{o}es e Comunica\c{c}\~{o}es (Brazil), and Korea Astronomy and Space Science Institute (Republic of Korea). This work was enabled by observations made from the Gemini North telescope, located within the Maunakea Science Reserve and adjacent to the summit of Maunakea. We are grateful for the privilege of observing the Universe from a place that is unique in both its astronomical quality and its cultural significance.

This work has made use of data from the European Space Agency (ESA) mission {\it Gaia} (\url{https://www.cosmos.esa.int/gaia}), processed by the {\it Gaia} Data Processing and Analysis Consortium (DPAC, \url{https://www.cosmos.esa.int/web/gaia/dpac/consortium}). Funding for the DPAC has been provided by national institutions, in particular the institutions participating in the {\it Gaia} Multilateral Agreement. This publication makes use of data products from the Wide-field Infrared Survey Explorer, which is a joint project of the University of California, Los Angeles, and the Jet Propulsion Laboratory/California Institute of Technology, funded by the National Aeronautics and Space Administration. This publication makes use of data products from the Two Micron All Sky Survey, which is a joint project of the University of Massachusetts and the Infrared Processing and Analysis Center/California Institute of Technology, funded by the National Aeronautics and Space Administration and the National Science Foundation. This research has made use of NASA's Astrophysics Data System. This research has made use of the VizieR catalogue access tool, CDS, Strasbourg, France. 

\end{acknowledgments}

\facilities{Sloan, 
Gemini:North, 
\textit{Swift}, 
FLWO:2MASS, 
\textit{WISE}, 
\textit{Gaia}, 
\textit{TESS}
}

\software{
\texttt{numpy} \citep{Harris2020},
\texttt{scipy} \citep{Virtanen2020},
\texttt{astropy} \citep{Robitaille2013,Price-Whelan2018},
\texttt{matplotlib} \citep{mpl},
\texttt{WD\_models} (\url{https://github.com/SihaoCheng/WD\_models}),
\texttt{wdtools} \citep{Chandra2020a,wdtools},
\texttt{lmfit} \citep{Newville2014},
\texttt{emcee} \citep{Foreman-Mackey2013,Foreman-Mackey2019},
\texttt{lightkurve} \citep{LightkurveCollaboration2018}
}

\bibliography{library}
\bibliographystyle{aasjournal}

\appendix

\section{MCMC Posteriors}

\restartappendixnumbering

\setcounter{figure}{0}

\begin{figure*}[!htb]
    \centering
    \includegraphics[width=\textwidth]{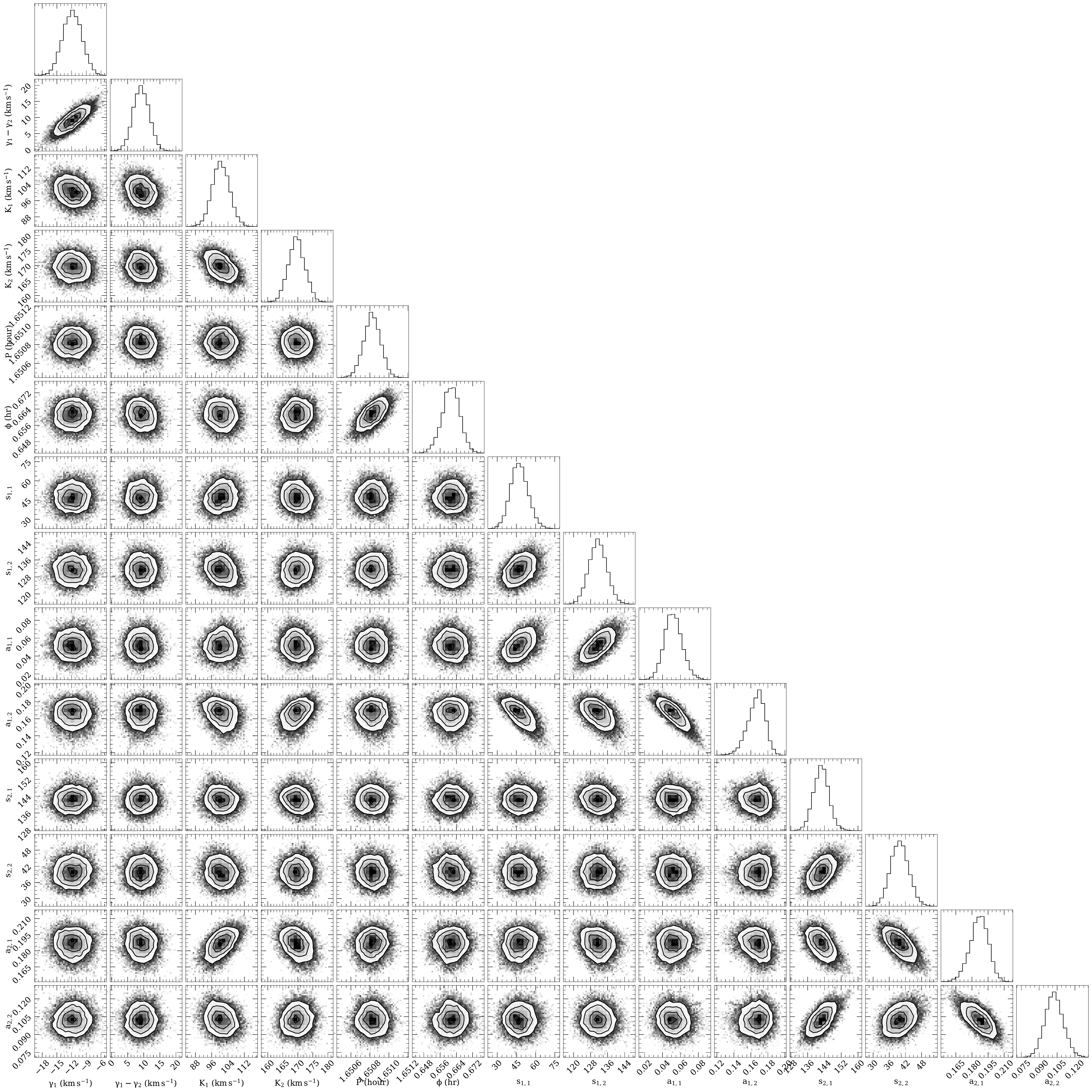}
    \caption{Corner plot of the posterior distributions of orbital parameters fitted to the time-resolved Gemini H$\alpha$ spectra. }
    \label{fig:halpha_corner}
\end{figure*}

\begin{figure*}
    \centering
    \includegraphics[width=\textwidth]{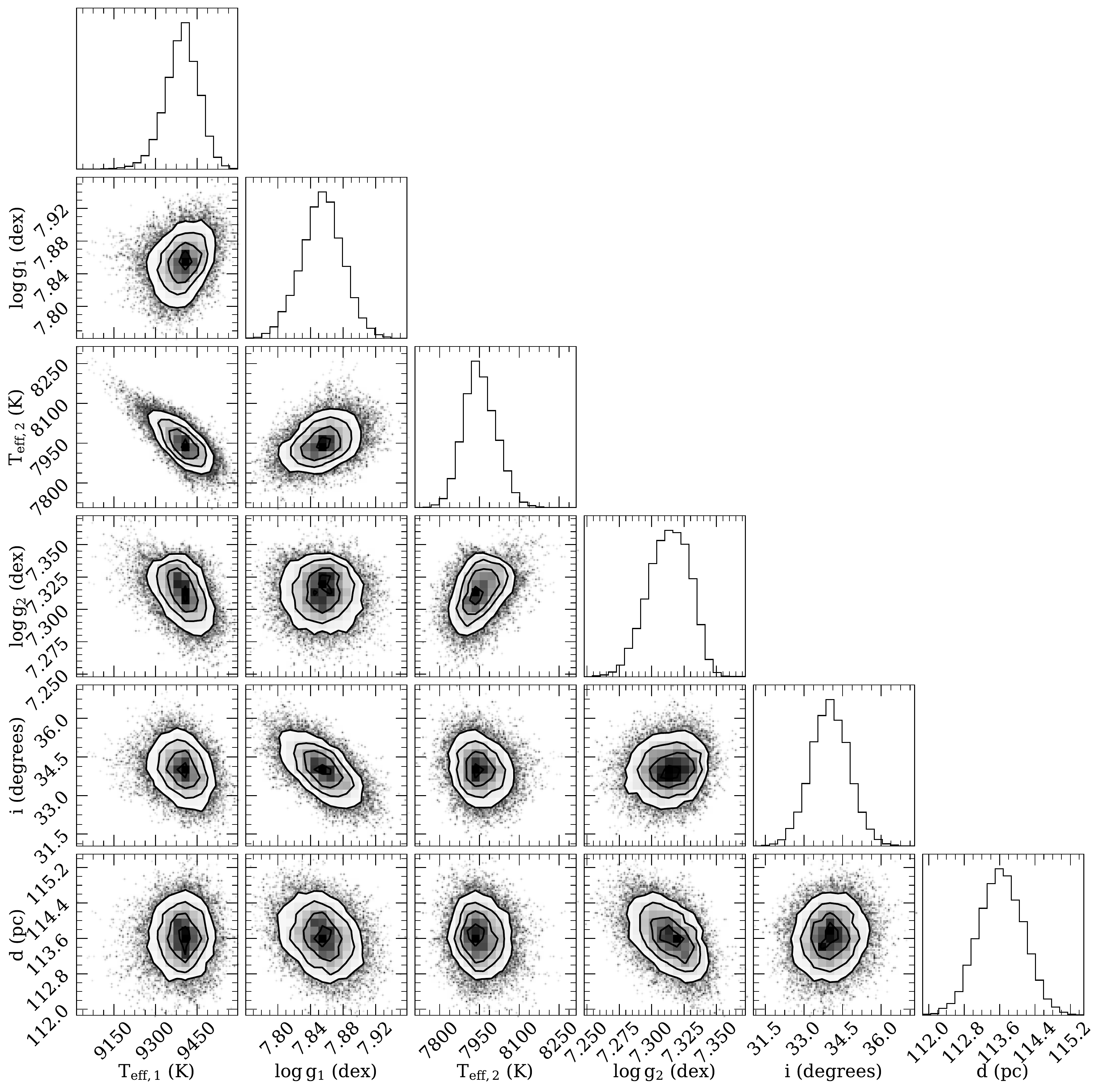}
    \caption{Corner plot of the posterior distributions of stellar parameters fitted simultaneously to the broadband photometry, SDSS-V spectrum, and Keplerian constraints.}
    \label{fig:phot_corner}
\end{figure*}

\end{document}